\documentclass[twoside]{article}
\usepackage{epsfig,fleqn,espcrc2}

\usepackage{graphicx}
\usepackage[figuresright]{rotating}

\newcommand{\AmS}{{\protect\the\textfont2
  A\kern-.1667em\lower.5ex\hbox{M}\kern-.125emS}}

\title{A smoother approach to scaling by suppressing monopoles and
vortices}

\author{Rajiv V. Gavai\address{Department of Theoretical Physics, Tata
                Institute of Fundamental Research, \\
                Homi Bhabha Road, Mumbai 400 005, India}%
        \thanks{E-mail: gavai@tifr.res.in} }
       
\begin{document}

\begin{abstract}
Suppressing monopoles and vortices by introducing large chemical
potentials for them in the Wilson action for the $SU(2)$ lattice
gauge theory, we study the nature of the deconfinement phase transition 
on $N_\sigma^3 \times N_\tau$ lattices for $N_\tau =$4,
5, 6 and 8 and $N_\sigma =$ 8--16. Using finite size scaling theory, we
obtain $\omega \equiv \gamma/\nu = 1.93 \pm 0.03$ for $N_\tau =$ 4,
in excellent agreement with universality. The critical couplings for
$N_\tau=$ 4, 5, 6 and 8 lattices exhibit large shifts towards the strong
coupling region when compared with the usual Wilson action, and 
suggest a lot smoother approach to scaling.
\vspace{1pc}
\end{abstract}

\maketitle

\section{INTRODUCTION}
Universality of lattice gauge theories implies that actions which differ
merely by irrelevant terms in the parlance of the renormalization group
lead to the same continuum quantum field theory.  While many numerical
simulations have been performed for the Wilson action for pure
gauge theories, other choices, some motivated by the desire to find a
smoother continuum limit, have also been employed. This usually has lead to
results in accord with universality.  Investigations of the deconfinement
phase transition for mixed actions, obtained by extending the  Wilson
action by addition  of an adjoint coupling term, showed \cite{us,Ste},
however, surprising challenges to the above notion of universality.
Several finite temperature investigations have established the presence of a
second order deconfinement phase transition for the Wilson action. Its
critical exponents have been shown \cite{EngFin} to be in very good agreement
with those of the three dimensional Ising model, as conjectured by
Svetitsky and Yaffe \cite{SveYaf}.  This verification of the universality
conjecture strengthened our analytical understanding of the
deconfinement phase transition which, however, came under a shadow of
doubt by the results for the mixed actions.  For two different actions,
it was found \cite{us,Ste} on a range of temporal lattice sizes that  
for large adjoint couplings a] the deconfinement transition became 
definitely of first order and b] no other separate bulk transition existed.
In fact, simulations on large symmetric lattices even suggested \cite{self} 
a lack of a bulk phase transition at an adjoint coupling where a first order 
deconfinement transition for a lattice of temporal size four was clearly seen.

Recently it was shown \cite{GavMan,DatGav} that suppression of some
lattice artifacts such as $Z_2$ monopoles and vortices do restore 
universality for one of the two actions: no first order deconfinement
transition was found in the entire coupling plane in that case.  Here we
address this question for the other action in the same manner and find
that one gains an additional bonus.  The approach to scaling seems to
become smoother than that for the original Wilson action.  

\section{ACTION AND RESULTS}

Both actions \cite{BhaCre,CaHaSc} have an adjoint coupling $\beta_A$ in
addition to the usual Wilson coupling $\beta$. The $\beta$ = 0 axis in
each case describes an $SO(3)$ model which has a first order phase
transition.  At $\beta_A$ = $\infty$, the theories reduce to a $Z_2$
lattice gauge theory with again a first order phase transition at
$\beta^{\rm crit} \approx$ 0.44.  The first order transitions extend
into the ($\beta$, $\beta_A$) coupling plane, meet at a triple point,
and continue as a single line of first order transitions which ends at
an apparently critical point, say, D in both cases.  The proximity of D
to the $\beta_A = 0$ line has commonly been held responsible for the
abrupt change from the strong coupling region to the scaling region for
the Wilson action.  Following Refs. \cite{GavMan,DatGav}, one can also
suppress $Z_2$ monopoles and $Z_2$ electric vortices for the Bhanot-Creutz 
action with suitable chemical potentials to restore universality
in its entire coupling plane. Using ${\rm sign (Tr}_F U_P)$ to identify
the $Z_2$ monopoles and vortices, one can define the corresponding action
for such suppressions as below:
\vspace{-.1cm}
\begin{eqnarray}
&&S_{BC,S} = \sum_P \beta \left( 1- {{\rm Tr}_F U_P \over 2}
\right) +  \nonumber \\
&&\sum_P \beta_A \left( 1 - {{\rm Tr}_A U_P \over 3} \right) 
+ \lambda_M \sum_c (1 - \sigma_c )\nonumber \\
&&~~~~~~~~~~~~~+ \lambda_E \sum_l ( 1 - \sigma_l )~,~
\label{bcs}~~~~~~ 
\end{eqnarray}
where $\sigma_c = \prod_{p \in \partial c} {\rm sign (Tr}_F U_P)$ and
$\sigma_l = \prod_{p \in \hat \partial l} {\rm sign (Tr}_F U_P)$.  It is
clear that the above theory flows to the same critical fixed point in
the continuum limit for all $\beta_A$, $\lambda_M$ and $\lambda_E$ and
has the same scaling behavior near the critical point.  It has to be
stressed though that universality has to be tested afresh for
eq.(\ref{bcs}) to be sure that the above naive argument about the
$\lambda_M$ and $\lambda_E$ terms being irrelevant is correct. This is
what we do in the following by determining a critical index of the
deconfinement phase transition. We then check whether the passage to
scaling is affected by studying the deconfinement transition as a
function of the temporal lattice size.  We chose $\beta_A$ =0,
$\lambda_M$ = 1 and $\lambda_E$ = 5 throughout.

We studied the deconfinement phase transition on $N_\sigma^3 \times
N_\tau$ lattices by monitoring its order parameter and the corresponding
susceptibility for $N_\tau =$ 4, 5, 6 and 8 and $N_\sigma =$ 8, 10, 12,
14, 15, and 16.  We used the simple Metropolis algorithm and tuned it to
have an acceptance rate $\sim 40$ \%.  The expectation values of the
observables were recorded every 20 iterations to reduce the
autocorrelations.  Errors were determined by correcting for the
autocorrelations and also by the jack knife method. The observables
monitored were the average plaquette, P, and the absolute value of the
average of the deconfinement order parameter $|L|$.  We also monitored
the susceptibilities for both $|L|$ and P: $\chi_{|L|}
=N_\sigma^3(\langle L^2 \rangle - \langle |L| \rangle ^2) $ and $
\chi_P =6 N_\sigma^3 N_\tau(\langle P^2 \rangle - \langle P
\rangle^2)$.

\subsection{$N_\tau = 4 $}
 \begin{table}
\caption{The values of $\beta$ at which long simulations were performed 
on $N_\sigma^3 \times 4$ lattices, $\beta_c$ and the height of 
the $|L|$-susceptibility peak, $\chi^{\rm max}_{|L|}$.}
\label{table:1}
\medskip
\begin{tabular}{@{}llll}
\hline
$N_\sigma$~~~ &~~~~ $\beta$~~~~  &~~~~  $\beta_{c,N_\sigma}$~~~~  & ~~~~ $\chi_{|L|}^{\rm max}$~~~~ \\
\hline
      8         &  ~~1.37     &  ~~1.366(7)     &    ~~~~~9.34(7)          \\
     10         &  ~~1.344    &  ~~1.360(5)     &    ~~~~14.34(11)         \\
     12         &  ~~1.331    &  ~~1.345(2)     &    ~~~~20.44(29)         \\
     14         &  ~~1.34     &  ~~1.343(2)     &    ~~~~27.48(64)         \\
\hline
\end{tabular} 
\vspace{-0.5cm}
\end{table}

The deconfinement phase transition on $N_\tau = 4 $ lattices was studied
by first making short hysteresis runs on the smallest lattice to look
for abrupt or sharp changes in the order parameter $ \langle |L|
\rangle$. In the region of its strong variation, longer runs were made
to check whether the $|L|$-susceptibility exhibits a peak.
Histogramming technique was used to extrapolate to nearby couplings for
doing this.  A fresh run was made at the $\chi_{|L|}$ peak position and
the process repeated until the input coupling for the run was fairly
close to the output peak position of the susceptibility.  The same
procedure was used for the bigger lattices also  but by starting from
the $\beta_c$ of the smaller lattice.  Typically 2-4 million Monte Carlo
iterations were used to estimate the magnitude of the peak height and
the peak location for each $N_\sigma$.   Table 1 lists our final results
for all the $N_\sigma$ used.  Fitting the peak heights in Table 1 to $A
N_\sigma^\omega$, we obtained $ \omega = 1.93 \pm 0.03$, in excellent
agreement with the values for both the standard Wilson
action \cite{EngFin} and the 3-dimensional Ising model. 

\subsection{$N_\tau = 5 $, 6 and 8}

\begin{table}
\caption{Same as Table 1 but for on $N_\sigma^3 \times 5$ lattices.}
\label{table:2}
\medskip
\begin{tabular}{@{}lllll}
\hline
$N_\sigma$ & $\beta$  &  $\beta_{c,N_\sigma}$  &  $\chi_{|L|}^{\rm max}$& $\chi^{\rm max}_{\rm predicted}$ \\

\hline
10         &  1.545    &  1.570(5)     &    13.17(17)&    --      \\
15         &  1.545    &  1.558(2)   &  29.57(77)& 28.78(41) \\
\hline
\end{tabular} 
\vspace{-0.5cm}
\end{table}

\begin{table}
\caption{Same as Table 1 but for on $N_\sigma^3 \times 6$ lattices.}
\label{table:3}
\medskip
\begin{tabular}{@{}lllll}
\hline
$N_\sigma$ & $\beta$  &  $\beta_{c,N_\sigma}$ & $\chi_{|L|}^{\rm max}$& $\chi^{\rm max}_{\rm predicted}$ \\
\hline
12         &  1.75     &  1.735(5)     & 16.34(45) &   --      \\
15         &  1.70     &  1.702(2) &  24.39(41)&  25.13(70)  \\
\hline
\end{tabular} 
\vspace{-0.5cm}
\end{table}

\begin{figure}[htbp]\begin{center}
\epsfig{height=7.5cm,width=6cm,file=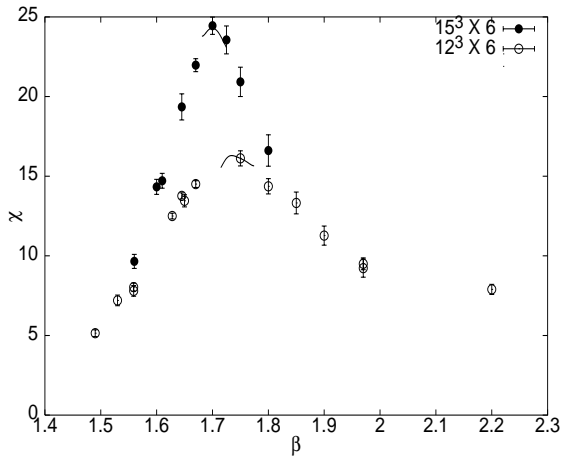,angle=270}
\vspace{-0.7cm}
\caption{The $|L|$-susceptibility as a function of $\beta$ on lattices
with $N_\tau$ = 6. The continuous lines are extrapolations using the 
histogramming technique.}
\vspace{-0.7cm}
\label{fg.chi6}\end{center}\end{figure}

For larger $N_\tau$, we 
used many longer runs in the region of strong variation of $\langle |L|
\rangle$ to obtain the susceptibility directly and used the histogramming only
for the finer determination of the critical coupling.  
Our results for $\langle |L| \rangle$ as a function of $\beta$ clearly 
show a deconfinement phase transition for $N_\tau$ = 5, 6 and 8.
This is also evident in the $\chi_{|L|}$ determinations, shown in 
Fig. \ref{fg.chi6} for $N_\tau = 6$.  
Tables 2 and 3 list the estimated maxima of $\chi_{|L|}$ for $N_\tau$ =
5 and 6 for two different spatial volumes along with the corresponding
peak locations.  Using our value for $\omega$, and the peak height for
the smaller spatial volume, the $\chi^{\rm max}$ on the bigger lattice
can be predicted.  These predictions are listed in the respective tables
in the last column and can be seen to be in very good agreement with the
direct Monte Carlo determinations.  We also determined $\omega$ from the
peak heights in Tables 2 and 3 and found it to be in good agreement with
the $N_\tau=4$ value.  Not only is the universality of the deconfinement
phase transition thus verified on three different temporal lattice
sizes, but it also confirms that the same {\it physical} phase
transition is being simulated on them, thus approaching the continuum
limit of $a \to 0$ in a progressive manner by keeping the transition
temperature $T_c$ constant in physical units.

\subsection{Scaling of $T_c$}

\begin{figure}[htbp]\begin{center}
\epsfig{height=7.5cm,width=6cm,file=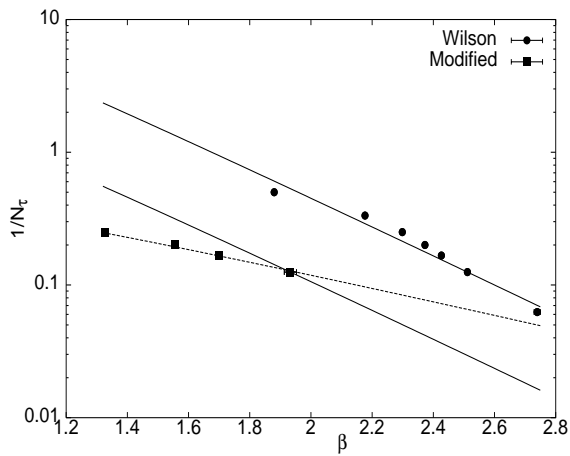,angle=270}
\vspace{-0.7cm}
\caption{$1/N_\tau$ as a function of $\beta_c$.  The full lines depict
the 2-loop asymptotic scaling relation,  while the dashed line denotes
eq.(\protect{\ref{rgs}}), normalized at $N_\tau$ = 8 in all cases. }
\vspace{-0.7cm}
\label{fg.scale}\end{center}\end{figure}

Fig. \ref{fg.scale} shows $aT_c = N_\tau^{-1}$ as a function of the
corresponding critical $\beta$ for both our simulations (squares) with
suppression of monopoles and vortices and the standard Wilson action
(circles). The latter are taken from the compilation of Ref.
\cite{Urs2}.  The dashed line describes a `phenomenological' scaling
equation,
\begin{equation}
 aT_c = {1 \over N_\tau} \propto \left({4b_0 \over \beta }\right)^{-b_1/b_0^2} 
\exp \left(- {\beta \over 4b_0}\right)~~,~~
\label{rgs}
\end{equation}
which is similar to the usual scaling relation, shown in Fig. 
\ref{fg.scale} by full lines, but with twice the exponent. 
All lines were normalized to pass through the $N_\tau$= 8 data points.  

One sees deviations from asymptotic scaling for both the Wilson action and 
our action with suppression of monopoles and vortices.  The deviations
for the same range of $N_\tau$ seem larger for our action but then one
is also considerably deeper in the strong coupling region of the
Wilson action where one {\it a priori} would not have even expected any
scaling behaviour.  As the agreement of our results with the dashed line 
of eq.(\ref{rgs}) in Fig. \ref{fg.scale} shows, scaling may hold in 
this region of $\beta$ for the suppressed action, since 
the relation between $a$ and $\beta$ in this region is similar
to the asymptotic scaling relation, differing only in the exponent which 
will cancel in dimensionless ratios of physical quantities.
It is clear that as $\beta \to \infty$, the difference between
the two actions must vanish. The data in Fig. \ref{fg.scale} do show such a 
trend although the limiting point is still not reached by $N_\tau$ = 8. 
It seems likely though that the trend of evenly spaced transition points
for our action will continue and the dashed line traced by its transition
points will merge with the Wilson action by $N_\tau \sim 25$ or so. 
If this were to be so, a much smoother approach to
continuum limit is to be expected after the suppression of monopoles
and vortices. In particular, one expects that dimensionless ratios of
physical quantities at the deconfinement phase transition couplings
should be constant, already from $\beta \sim 1.33 $, which is the
transition point for the $N_\tau$ = 4.

\section{SUMMARY}

The phase diagram of the Bhanot-Cruetz \cite{BhaCre} action in the
fundamental and adjoint couplings has been crucial in understanding many
properties of the $SU(2)$ lattice gauge theory, like the cross-over to the
scaling region from the strong coupling region.  Adding extra irrelevant
terms to the action one obtains a modified action (\ref{bcs}) in which 
monopoles and vortices can be suppressed by setting the additional couplings 
to large values.  Our finite size scaling analysis of $|L|$-susceptibility for 
$\beta_A$ = 0, $\lambda_M$ = 1 and $\lambda_E$ = 5, yielded 1.93
$\pm$ 0.03 for the critical exponent $\omega \equiv \gamma/\nu$ for
lattices with $N_\tau$ = 4,  thus verifying the naive universality. 
However, as a result of the suppression, the critical coupling is
shifted by about unity compared to the Wilson case.  Our results on
$N_\tau$ = 5 and 6 also yielded similar values for $\omega$ albeit with
larger errors, confirming that the same physical transition was being
studied this way as a function of the lattice cut-off, $a$.  While the data
for $N_\tau^{-1} = aT_c$ was found to vary slower with $\beta$ than expected 
from the asymptotic scaling relation for $N_\tau$ = 4--8, they 
did obey a similar relation with a factor of two larger exponent. A
straightforward extrapolation suggests that the results from the modified
action may merge with those of Wilson action for large $N_\tau$ ( of
about $\sim$ 25).  Thus the suppression appears to make the transition from 
strong coupling to the scaling region much smoother than that for
the unsuppressed Wilson action, allowing us to simulate the theory at
smaller $\beta$.  It will be interesting to see if dimensionless ratios
of physical quantities such as glueball masses or string tension with
$T_c$ are constant in the range of critical couplings explored here.
It will also be interesting to study such suppression in $SU(3)$, and indeed 
$SU(N)$ lattice gauge theories, since their phase diagrams are
similar and the same mechanism is expected to work for them as well.

\end{document}